\documentclass[aps,showpacs]{revtex4}

\usepackage{float}		      
\usepackage{dsfont}
\usepackage{graphicx}            
\usepackage{float}
\usepackage[T1]{fontenc}
\usepackage[latin1]{inputenc}
\usepackage{amssymb}
\include{epsf}
\epsfverbosetrue

\newcommand{\ket}[1]{\ensuremath{|#1\rangle}}			
\newcommand{\bra}[1]{\ensuremath{\langle#1|}}			
\newcommand{\iprod}[2]{\ensuremath{\langle#1|#2\rangle}}	
\newcommand{\oprod}[2]{\ensuremath{|#1\rangle\langle#2|}}	
\newcommand{\Tr}{\ensuremath{{\rm Tr}}}				
\newcommand{\bea}{\begin{eqnarray}}
\newcommand{\eea}{\end{eqnarray}}
\begin{document}
\title{Geometric measures of entanglement and the Schmidt decomposition}
\author{M.E.~Carrington}
\email{carrington@brandonu.ca}
\affiliation{Physics Department, Brandon University,
Brandon, Manitoba, R7A 6A9 Canada\\
and Winnipeg Institute for Theoretical Physics,
Winnipeg, Manitoba, Canada}
\author{R.~Kobes}
\email{r.kobes@uwinnipeg.ca}
\author{G.~Kunstatter}
\email{g.kunstatter@uwinnipeg.ca}
\author{D. ~Ostapchuk}
\email{davidostapchuk@alumni.uwaterloo.ca}
\affiliation{Physics Department, University of Winnipeg, Winnipeg, MB, R3B 2E9, Canada\\
and Winnipeg Institute for Theoretical Physics,
Winnipeg, Manitoba, Canada}
\author{G.~Passante}
\email{glpassan@iqc.ca}
\affiliation{Institute for Quantum Computing, University of Waterloo, Waterloo, Ont., N2L 3G1, Canada}
\begin{abstract}
In the standard geometric approach, the entanglement of a pure state is 
$\sin^2\theta$, where $\theta$ is the angle between the entangled state and the 
closest separable 
state of products of normalised qubit states. We consider here a generalisation of
this notion by considering separable states that consist of 
products of unnormalised states of different dimension. The 
distance between the target entangled state and the closest unnormalised product state can be interpreted as a measure of the entanglement of the target state. 
The components of the closest product state and its norm have
an interpretation in terms of, respectively, the eigenvectors and eigenvalues 
of the reduced density matrices arising in the Schmidt decomposition of 
the state vector.
For several cases where the target state has a large degree of symmetry, we  solve the system of equations analytically, and look specifically at the limit where the number of qubits is large.
\end{abstract}
\pacs{03.65.Ud, 03.67.Mn}
\maketitle

\section{Introduction}
\label{introSECT}
With recognition of its role as a resource in quantum computing \cite{nielsen},
the nature of entanglement in quantum systems
is a problem of much current interest \cite{review1,review2,review3}. 
Of particular importance is
a quantitative measure of entanglement \cite{quantify}. Two of the more commonly used measures
are the von Neumann entropy, which is based on reduced density matrices \cite{review1},
and a geometric measure, which is based on the distance to the nearest product
state \cite{g1,g2,g3,g4,witness}. In this paper we introduce a geometric measure of entanglement based on the distance between an 
{\it unnormalised} product state and a target entangled state. 
The norm of the closest product state
can be related to both the distance and angle
between the product and target states.
This result motivates the interpretation of the distance to the closest product state as a measure of the entanglement of the initial state.

\par We begin by defining our notation. We consider a system of $q$ qubits. The dimension of the corresponding Hilbert space is $n=2^q$. We will decompose the system into a set of subsystems. The subsystems are labelled $A,B,C,\cdots$ They have dimension $u,v,w,\cdots$ such that $n=u\cdot v\cdot w\cdots$. An arbitrary set of basis states of system $A$ is labelled $|i\rangle$, the basis states of $B$ are $|j\rangle$, the basis states of $C$ are $|k\rangle$, etc. Using this notation we write:
\bea
\ket{A}=\sum_{i=0}^{u-1} a_i\ket{i},~~\ket{B}=\sum_{j=0}^{v-1} b_i\ket{j},~~\ket{C}=\sum_{k=0}^{w-1} c_i\ket{k},~~\dots
\eea
We consider an arbitrary normalised entangled pure state and write its wave-function:
\bea
\label{psi}
\ket{\psi} = \sum_{i=0}^{u-1}\sum_{j=0}^{v-1}\sum_{k=0}^{w-1}\cdots \chi_{ijk\cdots} \ket{i}\otimes \ket{j}\otimes \ket{k}\cdots~,~~~\iprod{\psi}{\psi}=1\,.
\eea
In this paper we introduce a new geometric measure of the entanglement of this state. The paper is organised as follows. In section \ref{distSECT} we introduce our geometric measure of entanglement. In section \ref{schSECT} we show that, for a given entangled state, a connection can be established between the components and norm 
of the closest product state, and the basis states and eigenvalues of
the Schmidt decomposition of the entangled state.  In section \ref{symSECT} we study some general symmetries of our measure. In section \ref{exactSECT} we derive some exact solutions for cases where the target state $\ket{\psi}$ has a large degree of symmetry, and in section \ref{concSECT} we present our conclusions.

\section{Optimum Euclidean Distance}
\label{distSECT}
In this section we introduce a new geometric measure of entanglement. 
We look at the distance between the pure entangled state (\ref{psi}) and an arbitrary unnormalised product state. Extremizing this distance allows us to identify the closest product state. The distance between the state $\ket{\psi}$ and this closest product state is our geometric measure of the entanglement of $\ket{\psi}$. 

We consider the product state:
\begin{equation}
\label{phi}
 \ket{\phi} = \ket{A}\otimes\ket{B} \otimes \ket{C}\otimes\ldots
   = \sum_{i=0}^{u-1} a_i\ket{i} \otimes 
   \sum_{j=0}^{v-1} b_j\ket{j} \otimes
   \sum_{k=0}^{w-1} c_j\ket{k} \otimes\ldots
\end{equation}
The state $\ket{\phi}$ is not assumed to
be normalised:
\bea
\label{norm}
\iprod{\phi}{\phi} = N_AN_BN_C\ldots; ~~
N_A =\iprod{A}{A}=\displaystyle \sum_{i = 0}^{u-1}a_i^*a_i,~~
N_B =\iprod{B}{B} = \displaystyle \sum_{j = 0}^{v-1}b_j^*b_j,~~\cdots
\eea
The distance between the states $\ket{\psi}$ and $\ket{\phi}$ is:
\bea
\label{dist}
D^2 &&= \iprod{\phi - \psi}{\phi - \psi} \,,\nonumber\\
&& = 1 - \iprod{\phi}{\psi}-\iprod{\psi}{\phi}+N_A N_BN_C\ldots\nonumber\\
&&
= \left( \sum_{i = 0}^{u-1}\sum_{j = 0}^{v-1}\sum_{k = 0}^{w-1}\cdots \right)
\left(a_i^*b_j^*c_k^*\ldots - \chi_{ijk\cdots}^*\right)
\left(a_ib_jc_k\ldots - \chi_{ijk\cdots}\right)\,.
\eea
We extremize this distance with respect to the coordinates of 
$\ket{A}, \ket{B}, \ldots$
\begin{eqnarray}
\frac{\partial D^2}{\partial a_i} = 0 
&\Rightarrow& a_i^* N_BN_C\ldots =
\left(\sum_{j=0}^{v-1} \sum_{k=0}^{w-1}\cdots \right)
 b_jc_k\ldots\chi_{ijk\ldots}^*\nonumber\\
\frac{\partial D^2}{\partial b_j} = 0 
&\Rightarrow& b_j^* N_AN_C\ldots =
\left(\sum_{i=0}^{u-1} \sum_{k=0}^{w-1}\cdots \right)
a_ic_k\ldots \chi_{ijk\ldots}^*\nonumber\\
\frac{\partial D^2}{\partial c_k} = 0 
&\Rightarrow& c_k^* N_AN_B\ldots =
\left(\sum_{i=0}^{u-1} \sum_{j=0}^{v-1}\cdots \right)
a_ib_j\ldots \chi_{ijk\ldots}^*\nonumber\\
&\vdots&\label{opt}\end{eqnarray}
Rewriting Eq. (\ref{opt}), we have:
\begin{equation}
\label{csoln}
N_A N_BN_C\ldots =\left(\sum_{i=0}^{u-1}\sum_{j=0}^{v-1} \sum_{k=0}^{w-1}\cdots\right)
a_i b_jc_k\ldots\chi_{ijk\ldots}^* = \iprod{\psi}{\phi}\,.
\end{equation}
In exactly the same way we could extremize the distance in Eq. (\ref{dist}) with respect to the coordinates of $\bra{A}, \bra{B}, \ldots$ and obtain: $N_A N_BN_C\ldots = \iprod{\phi}{\psi}$.
Substituting these results into (\ref{dist}) we find that at the extrema the distance between $\ket{\psi}$ and $\ket{\phi}$ is:
\bea
D_C^2 = 1-N_AN_BN_C\ldots = 1-\cos^2\theta_C\,,
\label{cdistance}
\eea
where we have defined the critical angle $\theta_C$ as the angle between $\ket{\psi}$ and $\ket{\phi}$ at the extrema: 
\begin{equation}
\cos\theta_C = \left.
\frac{ \iprod{\psi}{\phi}}{\sqrt{\iprod{\phi}{\phi}}\sqrt{\iprod{\psi}{\psi}}}
\right|_{\rm critical} =
\sqrt{N_AN_BN_C\ldots}
\label{cangle}\end{equation}

\par

In order to demonstrate the consistency of these results, we look at the Cauchy--Schwartz inequality which requires:
\begin{equation}
\iprod{\phi - \psi}{\phi - \psi} \iprod{\phi}{\phi}  
\ge \left| \iprod{\phi - \psi}{\phi} \right|^2\,.
\label{cs}\end{equation}
Using (\ref{norm}) and (\ref{dist}) we have:
\begin{equation}
D^2\cdot(N_AN_BN_C\cdots)  \ge \left| (N_AN_BN_C\cdots) - \iprod{\psi}{\phi} \right|^2\,.
\end{equation}
Using (\ref{csoln}) and (\ref{cdistance}) we find that at the extrema
\begin{equation}
1-N_AN_BN_C\cdots \ge 0 ~~\Rightarrow ~~N_AN_BN_C\cdots \le 1\,.
\end{equation}
This inequality guarantees that the cosine of the critical angle (Eq. (\ref{cangle})) is a real number which satisfies $1>\cos\theta_C>-1$, and that the square of the critical distance (Eq. (\ref{cdistance})) is real and positive. \\

\par
We can compare the results in Eqs. (\ref{cdistance}), (\ref{cangle}) with those that would be obtained using a product of normalised states. We use $\iprod{A}{A}=\iprod{B}{B} =\cdots = 1$. We take the derivative of $D^2$ with respect to $a_i$ as before, but now we insert a Lagrange multiplier term of the form $\lambda \iprod{A}{A}$. This approach produces the same critical angle as before (Eq. (\ref{cangle})). 
The
corresponding minimal distance is:
\begin{equation}
D_N^2 = \iprod{\phi_N - \psi}{\phi_N - \psi} = 
D^2 + \left(1-\sqrt{\iprod{\phi}{\phi}}\right)^2
= 2(1-\cos\theta_C)\,,
\end{equation}
which can be compared with the result for unnormalised product states (Eq. (\ref{cdistance})). 
\begin{center}
\begin{figure}
\resizebox{\textwidth}{!} {
\includegraphics[width=4pt]{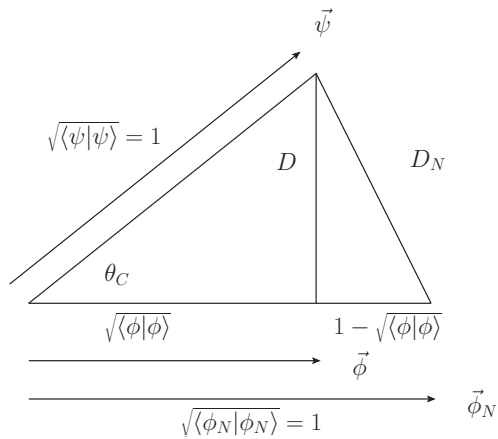}
}
\caption{Geometrical comparison of the entanglement measure using
unnormalised ($\vec \phi$) and normalised ($\vec \phi_N$) separable states.}
\label{geometry}
\end{figure}
\end{center}

\section{Connection to the Schmidt decomposition}
\label{schSECT}
The equations in (\ref{opt}) which 
determine the extremal points of the distance to the closest product
state are non--linear and must be solved numerically, except in special cases. One of the special cases for which a closed--form
solution exists occurs when the $n$--dimensional system is decomposed
into two subsystems. We consider a $u$-dimensional subsystem $A$, and a $v$-dimensional
subsystem $B$, such that $n = uv$. In this case the equations (\ref{opt}) 
decouple to yield:
\begin{eqnarray}
a_i N_AN_B &=& \sum_{i^\prime = 0}^{u-1}\sum_{j = 0}^{v-1} \chi_{ij}\chi_{i^\prime j}^*a_{i^\prime} \,,
\nonumber\\
b_j N_AN_B &=& \sum_{j^\prime = 0}^{v-1}\sum_{i = 0}^{u-1} \chi_{ij}\chi_{ij^\prime}^*b_{j^\prime}\,.
\label{extremal} 
\end{eqnarray}
Each of these equations can be solved for the product $N_AN_B$ and give, respectively:
\begin{eqnarray}
{\rm det} \left| N_AN_B \delta_{ii^\prime} - \sum_{j=0}^{v-1} \chi_{ij}\chi_{i^\prime j}^*
\right| &=& 0 \,,\nonumber\\
{\rm det} \left| N_AN_B \delta_{jj^\prime} - \sum_{i = 0}^{u-1} \chi_{ij}\chi_{ij^\prime}^*
\right| &=& 0\,.
\label{vn}\end{eqnarray}

These solutions can be related to the eigenvalues of the reduced density matrix.
We consider the $n$-dimensional state in Eq. (\ref{psi}) and write its wave-function and density matrix in the computational basis:
\bea
\ket{\psi}=\sum_{z=0}^{n-1}\chi_z\ket{z}\,,~~\rho = \sum_{z=0}^{n-1}\sum_{z^\prime=0}^{n-1}\chi_z\chi^*_{z^\prime}\ket{z}\bra{z}\,.
\eea
Decomposing the system into a $u$-dimensional subsystem $A$ and a $v$-dimensional subsystem $B$, we obtain:
\bea
\label{psi2}
&& \ket{\psi}  = \sum_{i=0}^{u-1}\sum_{j=0}^{v-1}\chi_{ij}\ket{i}\otimes\ket{j}\,,~~~\rho = \sum_{i=0}^{u-1} \sum_{i^\prime=0}^{u -1}\sum_{j=0}^{v-1}\sum_{j^\prime =0}^{v-1}\chi_{ij}\chi^{*}_{i^\prime j^\prime}\ket{i}\bra{i^\prime}\otimes \ket{j}\bra{j^\prime}\,.
\eea
Next we calculate the reduced density matrices. The reduced density matrix $\rho_A$ is obtained by tracing over the subsystem $B$, and the reduced density matrix $\rho_B$ is obtained by tracing over the subsystem $A$. The definitions are: 
\begin{eqnarray}
\Tr_A(\rho) &=& \sum_{t=0}^{u-1} \left( \bra{t} \otimes \mathds{1}_B\right) \rho
   \left( \ket{t} \otimes \mathds{1}_B\right) \,,\nonumber\\
\Tr_B(\rho) &=& \sum_{t=0}^{v-1} \left( \mathds{1}_A \otimes \bra{t} \right) \rho
   \left( \mathds{1}_A \otimes \ket{t} \right) \,,
\label{reduced}
\end{eqnarray}
where $\mathds{1}_A$ and $\mathds{1}_B$ are the identity matrices in the subspaces of
$A$ and $B$, respectively. We obtain:
\begin{eqnarray}
\rho_A = \Tr_B(\rho) &=&
\sum_{t=0}^{v-1} \sum_{i=0}^{u-1} \sum_{i^\prime=0}^{u-1} \sum_{j=0}^{v-1} \sum_{j^\prime=0}^{v-1} 
\chi_{ij} \chi_{i^\prime j^\prime}^*\iprod{t}{j}\iprod{j^\prime}{t} \oprod{i}{i^\prime} 
= \sum_{j=0}^{v-1} \sum_{i=0}^{u-1} \sum_{i^\prime=0}^{u-1} 
\chi_{ij} \chi_{i^\prime j}^* \oprod{i}{i^\prime}\,,
\end{eqnarray}
or, in terms of components:
\begin{equation}
\left(\rho_A\right)_{ii^\prime} = \sum_{j=0}^{v-1} \chi_{ij} \chi_{i^\prime j}^*\,. 
\label{rhoa}
\end{equation}
Similarly, the reduced density matrix
$\rho_B = \Tr_A(\rho)$ is obtained by tracing over the subsystem $A$ and can be written:
\begin{equation}
\left(\rho_B\right)_{jj^\prime}= \sum_{i=0}^{u-1} \chi_{ij} \chi_{ij^\prime}^* \,.
\label{rhob}
\end{equation}
Using 
Eqs. (\ref{rhoa}) and (\ref{rhob}) we can rewrite 
the extremal conditions of Eqs. (\ref{extremal}) in the form: 
\begin{eqnarray}
\sum_{i=0}^{u-1} \left(\rho_A\right)_{ii^\prime} a_{i^\prime} &=& N_A N_B a_i\,,\nonumber\\
\sum_{j=0}^{v-1} \left(\rho_B\right)_{jj^\prime} b_{j^\prime} &=& N_A N_B b_j\,.
\label{two}
\end{eqnarray}
Equation (\ref{two}) shows that $N_AN_B$ are the eigenvalues
 corresponding to the eigenvectors $a_i$ and $b_j$ of the reduced
density matrices $\rho_A$ and $\rho_B$. 

This result can be interpreted in terms of the Schmidt
decomposition as follows \cite{nielsen,s1,s2,s3,s4,s5}. Let us write the extrema conditions of
Eq.(\ref{opt}) as
\begin{eqnarray}
{\widehat a_i} \sigma &=& \sum_{j=0}^{v-1} \chi_{ij}\, {\widehat b_j}^*\nonumber\\
{\widehat b_j} \sigma &=& \sum_{i=0}^{u-1}  {\widehat a_i}^* \, \chi_{ij}
\end{eqnarray}
where ${\widehat a_i} = a_i / \sqrt{N_A}$ and  ${\widehat b_j} = b_j / \sqrt{N_B}$ are, respectively,
the left--singular and right--singular vectors corresponding to the singular values
$\sigma = \sqrt{N_AN_B}$
of the matrix $\chi$. The singular value decomposition of the matrix $\chi$ can then be written as
\begin{equation}
\chi = A\, \Sigma \,B^\dagger
\end{equation}
where the columns of the (unitary) matrices $A$ and $B$ are, respectively, the vectors ${\widehat a_i}$
and ${\widehat b_j}$, and $\Sigma$ is a diagonal matrix whose elements are the singular values
$\sigma$. This can be used to rewrite the state $\ket{\psi}$ in (\ref{psi2}) in terms of the Schmidt
decomposition involving a single summation:
\begin{equation}
\ket{\psi} = \sum_{k=0}^{ {\rm min}(u-1, v-1) } \sqrt{p_k} \ 
\ket{\alpha_k} \otimes \ket{\beta_k} \,,
\label{sd}\end{equation}
where the Schmidt coefficients $p_k$, which are identified with
the singular values $\sigma$, satisfy $\sum_k p_k = 1$, and the 
states $\ket{\alpha_k}$ and $\ket{\beta_k}$ are identified with, respectively,
the left--singular and right--singular vectors ${\widehat a_i}$
and ${\widehat b_j}$.
Calculating the corresponding reduced density matrices using (\ref{reduced}) we obtain:
\begin{eqnarray}
\label{six}
\rho_A &=& \Tr_B(\rho) = \sum_k p_k \ \oprod{\alpha_k}{\alpha_k} \,,\nonumber\\
\rho_B &=& \Tr_A(\rho) = \sum_k p_k \ \oprod{\beta_k}{\beta_k}\,,
\end{eqnarray}
from which one can see that $\ket{\alpha_k}$ are the eigenvectors of $\rho_A =\Tr_B(\rho)$
and $\ket{\beta_k}$ are the eigenvectors of $\rho_B = \Tr_A(\rho)$ 
with corresponding eigenvalues $p_k$.

We remark that the Schmidt decomposition gives rise to several other measurements of entanglement. The reduced density matrices $\rho_A$ and $\rho_B$ in (\ref{six}) have the same non-zero eigenvalues and, for a product
state, only one non--zero eigenvalue is present.  As a result, for a non-product state, the eigenvalues of the reduced density matrix can  be used to
quantify the degree of entanglement. One 
commonly used measure of entanglement is the von Neumann entropy:
\begin{equation}
S = -\Tr\left(\rho_A \ln \rho_A\right) = -\Tr\left(\rho_B \ln \rho_B\right)
= -\sum_k p_k \log_2 p_k\,.
\end{equation}

\par
It is interesting to consider
the particular case that the $n$--dimensional space is split into
the product of a single qubit space and another space of dimension
$u=n/2$. In this case,  
one of the equations in (\ref{vn}) will become a quadratic equation
for the product $N_AN_B$ with solutions:
\begin{equation}
N_AN_B \equiv \mu_\pm = \frac{1}{2}\left[ 1 \pm \sqrt{1-4C}\right]
, \qquad\qquad C = \sum_{j=1}^{u-1} \sum_{k=0}^{j-1} \left| \chi_{0j}\chi_{1k} - \chi_{1j}\chi_{0k}\right|^2\,.
\end{equation}
The Schmidt decomposition (Eq.~(\ref{sd})) becomes:
\begin{equation}
\label{psi3}
\ket{\psi} = \sqrt{\mu_+} \ \ket{\alpha_+} \otimes \ket{\beta_+} +
 \sqrt{\mu_-} \ \ket{\alpha_-} \otimes \ket{\beta_-}\,.
\end{equation}
Using Eq.~(\ref{cangle}) we can relate the eigenvalues $N_AN_B=\mu_\pm$ to the cosine of the angle
between $\ket{\psi}$ and the closest product state $\ket{\phi}$. 
Since $\mu_+$ is the larger of the
two eigenvalues we write $\cos\theta_{\rm max} \equiv \mu_+$ and Eq. (\ref{psi3}) becomes:
\begin{equation}
\ket{\psi} = \cos\theta_{\rm max} \ \ket{\alpha_+} \otimes \ket{\beta_+} +
 \sin\theta_{\rm max} \ \ket{\alpha_-} \otimes \ket{\beta_-}\,.
\end{equation}
If $\ket{\psi}$ was a product state, the reduced density matrix would have only one non-zero eigenvalue and 
we would have $\mu_- =\sin^2\theta_{\rm max} = 0$. 
This result is consistent with our geometric approach: from Eq. (\ref{cdistance}), $\sin^2\theta_{\rm max}=0$ corresponds to a zero minimal distance between the state $\ket{\psi}$ and the nearest product state, which means that $\ket{\psi}$ is itself a product state. \\

The
Schmidt decomposition can also be applied to multipartite pure states \cite{s4}. One starts with a state $\ket{\psi}$
and decomposes it into two subsystems: a single qubit system $A$,
and a subsystem ($BC\ldots Z$) containing all other qubits. Using a Schmidt decomposition we can write:
\begin{equation}
\ket{\psi} = \sum_{i_a} \sqrt{p_{i_a}^A} \ket{\psi_{i_a}^A} 
\otimes \ket{\psi_{i_a}^{BC\ldots Z}}\,.
\label{stage1}\end{equation}
One then decomposes $ \ket{\psi_{i_a}^{BC\ldots Z}} $ into two
subsystems: another single qubit system $B$, and a subsystem 
($CD\ldots Z$) containing all other qubits. Using a Schmidt decomposition we can write:
\begin{equation}
\ket{\psi_{i_a}^{BC\ldots Z}} = \sum_{i_b} \sqrt{p_{i_a;i_b}^B}
\ket{\psi_{i_a;i_b}^B} \otimes \ket{\psi_{i_a;i_b}^{CD\ldots Z}}\,.
\label{stage2}\end{equation}
This process is continued until the last two qubit spaces
$Y$ and $Z$ are reached, with the result:
\begin{equation}
\ket{\psi} = \sum_{i_a i_b \ldots i_{y} } 
\sqrt{p_{i_a}^A p_{i_a;i_b}^B \cdots 
 p_{i_a i_b \ldots i_x;i_{y} }^{Y}  } \ket{\psi_{i_a}^A} 
\otimes \ket{\psi_{i_a;i_b}^B} \otimes\cdots\otimes
 \ket{\psi_{i_a i_b \ldots i_x;i_{y} }^{Y} } \otimes
 \ket{\psi_{i_a i_b \ldots i_x;i_{y} }^{Z} }\,.
\end{equation}

Now we consider the geometric interpretation of this result. Consider the distance
between $\ket{\psi}$ 
and a state $\ket{\phi^{A;BC\ldots Z}} =  \ket{\phi^A} 
\otimes \ket{\phi^{BC\ldots Z}}$, where $\ket{\phi^A}$ is a single qubit state and $\ket{\phi^{BC\ldots Z}}$ is the state of the remaining qubits. This distance is: 
\begin{equation}
D_{A;BC\ldots Z}^2 = \big\langle \psi - \left[
\phi^A \otimes \phi^{BC\ldots Z} \right] \big| \psi - \left[
\phi^A \otimes \phi^{BC\ldots Z} \right] \big\rangle\,.
\end{equation}
Finding the extremal points of this distance
will result in a system of (linear) equations, as in Eq.~(\ref{vn}),
which determine the components of the state $\ket{\phi^{A;BC\ldots Z}}$.
There is a direct correspondence between the coefficients of the closest product state and
the basis states of the Schmidt decomposition (Eq. (\ref{stage1})),
and the norm of the closest product state is
related to the corresponding Schmidt coefficients.

We can then consider the
distance between the state $\ket{\phi^{BC\ldots Z}} $ and a state
$\ket{\phi^{B;CD\ldots Z}} =  \ket{\phi^B} 
\otimes \ket{\phi^{CD\ldots Z}}$:
\begin{equation}
D_{B;CD\ldots Z}^2 =\big\langle \phi^{BC\ldots Z} - \left[
\phi^B \otimes \phi^{CD\ldots Z} \right] \big| \phi^{BC\ldots Z} - \left[
\phi^B \otimes \phi^{CD\ldots Z} \right] \big\rangle\,,
\end{equation}
where $\ket{\phi^B}$ is a single qubit state and $\ket{\phi^{CD\ldots Z}}$ is the state of the remaining qubits. Extremizing this
distance will again result in a system of linear equations
determining the components of the state $\ket{\phi^{B;CD\ldots Z}}$. Once again, there is a direct correspondence between the coefficients of the closest product state and
the basis states of the Schmidt decomposition (Eq. (\ref{stage2})),
and the norm of the closest product state is
related to the corresponding Schmidt coefficients.

This process may be continued until the last two qubit states
$\ket{\phi^Y}$ and $\ket{\phi^Z}$ are reached. 
In analogy with Eq. (\ref{cangle}), we can define
the cosine of the critical angle $\theta_C$:
\begin{equation}
\cos\theta_C = \left. \frac{
 \iprod{\psi}{\phi} }
 { \sqrt{\iprod{\phi}{\phi}}\sqrt{\iprod{\psi}{\psi}}}
\right|_{\rm critical} \qquad\qquad \ldots\quad
\ket{\phi} = \ket{\phi^A} \otimes \ket{\phi^B} \otimes \cdots \ket{\phi^Z}\,.
\end{equation}
The quantity $\sin^2\theta_C = 1-\cos^2\theta_C$ is a measure of the entanglement of the original multipartite state. 

The advantage of this procedure is that it involves solving a series of linear equations, as compared to the approach of
Section \ref{distSECT}  which produces the non--linear equations of Eq.~(\ref{opt}). 
The
disadvantage is that the 
result depends on the order that the series of decompositions is made:
the sequence $ABC\ldots YZ$ described above will differ from the
sequence $BC\ldots ZA$. 
In Ref.~\cite{s4}, the entanglement is given by the minimal value obtained by looking at all permutations
of the possible orders of the decompositions.

\section{Symmetries}
\label{symSECT}
The equation that gives the distance between the target entangled state and the nearest product state (Eq. (\ref{cdistance})) is invariant under certain transformations 
of the parameters $\{a_i,b_i,\ldots\}$.
In order to see these symmetries explicitly we can reparameterize each set of coefficients. For the coefficients $a_i$ we write:
\begin{eqnarray}
a_i &=& A_i  e^{i \alpha_i^a}\,.
\end{eqnarray}
with similar equations for the coefficients $\{b_i,c_i,\ldots\}$.
Using generalised spherical coordinates in $n$ dimensions we can rewrite the set of real variables $A_i$ in terms of a magnitude $A$ and ($n-1$) angles $\theta_i$:
\begin{equation}
\displaystyle \left( \begin{array}{c}
A_1\\ A_2 \\ A_3 \\ \vdots \\ A_k \\ \vdots \\ A_{n-1} \\ A_n
\end{array} \right) = 
A \left( \begin{array}{c}
\cos\theta_1 \\ \sin\theta_1\cos\theta_2 \\ \sin\theta_1\sin\theta_2\cos\theta_3 \\ \vdots \\
\left( \Pi_{i=1}^{k-1} \sin\theta_i \right) \cos\theta_k \\ \vdots \\
\sin\theta_1\cdots\sin\theta_{n-2}\cos\theta_{n-1} \\
\sin\theta_1\cdots\sin\theta_{n-2}\sin\theta_{n-1}
\end{array} \right) \equiv Af_i(\theta^a)\,.
\end{equation}
We can also parameterise the phase angles $\alpha_i^a$ as:
\begin{equation}
e^{i\alpha_i} \left( \begin{array}{c}
 e^{i\alpha_1} \\ e^{i\alpha_2} \\ \vdots \\ e^{i\alpha_n} 
\end{array}  \right) =
e^{i\alpha_1}  \left( \begin{array}{c}
1 \\ e^{i(\alpha_2-\alpha_1)} \\ \vdots \\ e^{i(\alpha_n-\alpha_1)} 
\end{array}  \right) \equiv
e^{i\alpha_1}  \left( \begin{array}{c}
1 \\ e^{i\beta_2} \\ \vdots \\ e^{i\beta_n} 
\end{array}  \right) \equiv e^{i\alpha_1} e^{i\beta_i} \,,
\end{equation}
where $\beta_1 \equiv 0$.
Using this notation we write: 
\begin{eqnarray}
a_i &=& A e^{i\alpha_1^a}  f_i(\theta^a) e^{i \beta_i^a}\nonumber\\
b_i &=& B e^{i\alpha_1^b}  f_i(\theta^b) e^{i \beta_i^b}\nonumber\\
&&\vdots
\end{eqnarray}
We can use this parameterisation to rewrite the distance function (\ref{cdistance}). We make the definitions:
\bea
\label{three}
\iprod{\phi}{\phi}&&\equiv N^2 \,,\nonumber\\
\ket{\phi}&&\equiv N\ket{\hat\phi}\,,~~~\iprod{\hat\phi}{\hat\phi}=1\,.
\eea
Note that the first line in (\ref{three}) gives:
\bea
\iprod{\phi}{\phi}=A^2 B^2C^2\cdots=N_A N_B N_c\cdots = N^2\,.
\label{four}
\eea
Using these definitions the distance in Eq. (\ref{dist}) becomes:
\begin{equation}
\label{Dhat}
D^2 = N^2 - N \left[ \iprod{\psi}{ {\hat \phi} } + 
\iprod{ {\hat \phi} }{\psi} \right] + 1\,.
\end{equation}
Extremizing we obtain: 
\begin{eqnarray}
\frac{\partial D^2}{\partial N} = 0 &\Rightarrow&
 2N - \left[ \iprod{\psi}{ {\hat \phi} } + 
 \iprod{ {\hat \phi} }{\psi} \right] = 0 \,,\nonumber\\
&\Rightarrow& N = {\rm Re} \left[  \iprod{\psi}{ {\hat \phi} } \right ]\,,
\label{extrema}
\end{eqnarray}
which means that at the extrema:
\begin{equation}
\label{five}
D_C^2 = N^2 - N(2N) + 1 = 1-N^2\,.
\end{equation}
We note that Eqs. (\ref{four}) and (\ref{five}) are consistent with (\ref{cdistance}).

To make this result more clear, we can look explicitly at the dependence of the distance function on the overall phase of the coefficients of the product state. We define the overall phase angle $\delta = \alpha_1^a + \alpha_1^b+\ldots$ and obtain:
\begin{eqnarray}
\iprod{\phi}{\psi} &=&N \iprod{ {\hat \phi} }{\psi} =  N e^{-i\delta}
\sum_{i = 1}^{u}\sum_{j = 1}^{v} \cdots 
f_i(\theta^a)e^{-i\beta_i^a} f_j(\theta^b)e^{-i\beta_j^b} \ldots \chi_{ij\cdots}\,,  \nonumber\\
 \iprod{\psi}{\phi} &=&N \iprod{\psi}{ {\hat \phi} } =  N e^{i\delta}
\sum_{i = 1}^{u}\sum_{j = 1}^{v} \cdots 
f_i(\theta^a)e^{i\beta_i^a} f_j(\theta^b)e^{i\beta_j^b} \ldots \chi_{ij\cdots}^*\,, \label{phipsi}
\end{eqnarray}
which gives:
\begin{eqnarray}
\frac{\partial}{\partial \delta} \iprod{\psi}{\phi} &=& 
N \frac{\partial}{\partial \delta} \iprod{\psi}{{\hat \phi} } = iN  \iprod{\psi}{{\hat \phi} }\,,\nonumber\\
\frac{\partial}{\partial \delta} \iprod{\phi}{\psi} &=& 
N \frac{\partial}{\partial \delta} \iprod{{\hat \phi} }{\psi} =
-iN  \iprod{\psi}{{\hat \phi} }\,.
\end{eqnarray}
From Eq. (\ref{Dhat}) we have:
\begin{equation}
\frac{\partial D^2}{\partial \delta}
= -N\frac{\partial}{\partial \delta}
\left[ \iprod{\psi}{ {\hat \phi} } + 
\iprod{ {\hat \phi} }{\psi} \right]  = -iN \left[ \iprod{\psi}{ {\hat \phi} } -
\iprod{ {\hat \phi} }{\psi} \right] = 2N{\rm Im} \iprod{\psi}{ {\hat \phi} }\,,
\end{equation}
which means that at the extrema,
$ {\rm Im}  \iprod{\psi}{ {\hat \phi} } =0$. 
From Eq. (\ref{extrema}) we obtain that at the extrema:
\bea N = \iprod{\psi}{ {\hat \phi} } = \iprod{ {\hat \phi} }{\psi}\,.\eea 
This means that the critical angle defined by:
\begin{equation}
\cos{\hat \theta_C} = \left.
\frac{ \iprod{\psi}{{\hat \phi}}}{\sqrt{\iprod{{\hat \phi}} {{\hat \phi}}}\sqrt{\iprod{\psi}{\psi}}}
\right|_{\rm critical} = N\,,
\end{equation}
is the same critical angle as in  Eq. (\ref{cangle}).
\par

\section{Exact Solutions}
\label{exactSECT}

In this section we look at some states with a large degree of symmetry for which the equations (\ref{opt}) can be solved exactly. We consider a system of $q$ qubits and divide the Hilbert space of dimension $n=2^q$ into $q$ spaces, each of dimension 2. Using the notation of section \ref{introSECT} we have $n=u\cdot v\cdot w\cdots$ with $u=v=w=\cdots = 2$. The basis states in each single qubit system are:
\bea
\ket{0}=\left(\begin{array}{c}1\\0\end{array}\right)\,,~~\ket{1}=\left(\begin{array}{c}0\\1\end{array}\right)\,.
\eea
The product state in Eq. (\ref{phi}) becomes:
\bea
\ket{\phi}&& = (a_0\ket 0+a_1\ket 1) (b_0\ket 0+b_1\ket 1) (c_0\ket 0+c_1\ket 1)\cdots\nonumber\\
&& = (a_0 b_0 c_0\cdots )\ket{0,0,0,\cdots} + (a_0 b_0 c_1\cdots) \ket{0,0,1,\cdots}+ (a_0 b_1 c_1\cdots) \ket{0,1,1,\cdots} +\cdots
 \eea
 We look for a solution of the form:
\begin{eqnarray}
\label{randyRule}
a_0 &=& b_0 = \cdots = \alpha_0\,, \nonumber\\
a_1 &=& b_1 = \cdots = \alpha_1\,,
\end{eqnarray}
which means $N_A=N_B=\cdots\equiv N$.
\subsection*{Case 1}
Consider:
\bea
\ket{\psi} = \sqrt{p}\, \ket{0\,0\,\ldots 0} + \sqrt{1-p}\, \ket{1\,1\,\ldots 1}\,.
\eea
The 
only two non--zero components of $\psi_{ijk\cdots}$ are:
\begin{eqnarray}
\psi_{00\cdots 0} &=& \sqrt{p}\,, \nonumber\\
\psi_{11\cdots 1} &=& \sqrt{1-p} \,,
\end{eqnarray}
and Eq. (\ref{opt}) becomes:
\begin{eqnarray}
a_0 N_B N_C \cdots &=& \sqrt{p}\, b_0c_0\cdots\nonumber\\
a_1 N_B N_C \cdots &=& \sqrt{1-p}\, b_1c_1\cdots
\label{a1}\end{eqnarray}
Using Eq. (\ref{randyRule}), Eq. (\ref{a1}) becomes:
\begin{eqnarray}
\alpha_0 N^{q-1} &=&\sqrt{p}\, \alpha_0^{q-1}\,, \nonumber\\
\alpha_1 N^{q-1} &=& \sqrt{1-p}\, \alpha_1^{q-1}\,.
\label{a2}\end{eqnarray}
Solving this set of equations we obtain:
\bea
\alpha_1^2&&=  \frac{ N^{ 2(q-1)/(q-2)} }{ (1-p)^{1/(q-2)} }\,,\nonumber\\
\alpha_0^2&& = \frac{ (1-p)^{1/(q-2)} }{ p^{1/(q-2)} } \alpha_1^2\,,
\eea
which gives:
\begin{eqnarray}
N= \alpha_0^2 + \alpha_1^2 
 &=& \frac{ N^{ 2(q-1)/(q-2)} }{ (1-p)^{1/(q-2)} } 
  \left[
 1 +  \frac{ (1-p)^{1/(q-2)} }{ p^{1/(q-2)} } \right] \,.
\end{eqnarray}
Rearranging we obtain:
\begin{equation}
N^q = \frac{ p(1-p) } { \left[ 
p^{ 1/(q-2)} + (1-p)^{1/(q-2)} \right]^{q-2} }\,.
\end{equation}
\par
We can look at the large $q$ limit. 
Defining 
$A =  \left[ p^{ 1/(q-2)} + (1-p)^{1/(q-2)} \right]^{q-2}$ we have,
\begin{eqnarray}
\log A &=& (q-2) \log\left[ p^{ 1/(q-2)} + (1-p)^{1/(q-2)} \right] \,,\nonumber\\
&=& (q-2) \log(p^{ 1/(q-2)} ) + (q-2) \log\left[ 1+ \left( \frac{1-p}{p}\right)^{1/(q-2)} \right] \,,\nonumber\\
&\approx& \log p + (q-2)\log 2 + \frac{1}{2}\log\left( \frac{1-p}{p}\right)\,, \nonumber\\
&=& (q-2)\log 2 + \frac{1}{2}\log\left[ p(1-p)\right]\,, \nonumber\\
&=& \log\left[ 2^{q-2} \sqrt{p(1-p)} \right]\,,
\end{eqnarray}
which gives
\begin{equation}
\label{res1}
N^q \approx \frac{ \sqrt{p(1-p)}}{2^{q-2}}\,.
\end{equation}
\subsection*{Case 2}
Consider: 
\bea
\ket{\psi} =  \left[
\, \ket{1\,0\,\ldots 0} +\, \ket{0\,1\,\ldots 0} +\ldots + \ket{0\,0\,\ldots\,1}\right]/\sqrt{q}\,.
\eea
The 
only non--zero coefficients $\psi_{ijk\cdots}$ are:
\begin{equation}
\psi_{100\cdots 0} = \psi_{010\cdots 0}  = \psi_{001\cdots 0} = \cdots=\psi_{000\cdots 1} =
\frac{1}{\sqrt{q}}\,,
\end{equation}
and Eq. (\ref{opt}) becomes:
\begin{eqnarray}
a_0 N_B N_C \cdots &=& \frac{1}{\sqrt{q}} \left[ b_1c_0d_0\cdots + b_0c_1d_0\cdots+b_0c_0d_1\cdots+
\ldots\right]\nonumber\\
a_1 N_B N_C\cdots &=& \frac{1}{\sqrt{q}} \,b_0 c_0 d_0\cdots 
\label{b1}\end{eqnarray}
Using Eq. (\ref{randyRule}), Eq. (\ref{b1}) becomes:
\begin{eqnarray}
\alpha_0 N^{q-1} &=& \psi\, \alpha_1 \alpha_0^{q-2} (q-1)\,, \nonumber\\
\alpha_1 N^{q-1} &=& \psi\, \alpha_0^{q-1}\,.
\label{b2}\end{eqnarray}
Solving the set of equations in (\ref{b2}) we obtain:
\bea
\alpha_1^{q-2} &&= \frac{\sqrt{q} N^{q-1} }{(q-1)^{(q-1)/2} }\,,\nonumber\\
\alpha_0^2 &&= \alpha_1^2 (q-1)\,,
\label{b3}
\eea
which gives:
\begin{equation} 
N=\alpha_0^2 + \alpha_1^2 = \frac{ N^{ 2(q-1)/(q-2) } }
{\left( 1- \frac{1}{q} \right)^{ (q-1)/(q-2) } }\,.
\label{b4}\end{equation}
Rearranging we obtain:
\begin{equation}
\label{res2}
N^q = \left( 1-\frac{1}{q}\right)^{q-1}\,.
\end{equation}
\subsection*{Case 3}
Consider: 
\bea
\ket{\psi} =  \left[
\, \ket{1\,1\,0\,0\,\ldots 0} +\, \ket{0\,1\,1\,0\,\ldots 0}+\, \ket{0\,0\,1\,1\,\ldots 0} +\ldots + \ket{0\,0\,\ldots\,1\,1}+ \ket{1\,0\,\ldots\,0\,1}\right]/\sqrt{q}\,.
\eea 
The only non--zero coefficients $\psi_{ijk\cdots}$ are:
\begin{equation}
\psi_{11000\cdots 0} = \psi_{01100\cdots 0}  = \psi_{00110\cdots 0} =\psi_{00011\cdots 0}= \cdots=\psi_{00000\cdots 11} =\psi_{10000\cdots 01}=
\frac{1}{\sqrt{q}}\,.
\end{equation}
Using Eq. (\ref{randyRule}),  
Eq. (\ref{opt}) gives:
\begin{eqnarray}
 N^{q-1} &=& 2  \alpha_1^{q-2} \sqrt{q}\,, \nonumber\\
N^{q-1} &=& \sqrt{q}\alpha_0^2\alpha_1^{q-4}-\frac{2}{\sqrt{q}}\alpha_0^2\alpha_1^{q-4}\,.
\label{d2}\end{eqnarray}
Solving the equations in (\ref{d2}) we obtain:
\bea
\frac{\alpha_0}{\alpha_1} &&= \frac{\sqrt{2}}{\sqrt{q-2}}\,, \nonumber\\
\alpha_1&&=\frac{1}{2^{2-q}}\frac{1}{q^{2(q-2)}}N^{(q-1)/(q-2)}\,,
\label{d3}
\eea
which gives:
\begin{equation} 
N=\alpha_0^2 + \alpha_1^2 = \frac{1}{Q-2}2^{1/(2(2-q))}q^{1+1/(q-2)}N^{2(q-1)/(q-2)}\,.
\label{d4}\end{equation}
Rearranging we extract:
\begin{equation}
\label{res3}
N^q = 4(q-2)^{q-2}q^{1-q}\,.
\end{equation}

\subsection*{Case 4}
Consider the case where $\ket{\psi}$ has $q$ qubits consisting of all possible combinations
of $p$ entries of ``1'' and $q-p$ entries of ``0''. The normalisation is:
\begin{equation}
\psi^{-1} \equiv \sqrt {{ q \choose p}} = \sqrt{ \frac{ q!}{p!(q-p)! }}
\label{cnorm}\end{equation}
Using Eq. (\ref{randyRule}), Eq. (\ref{opt}) becomes:
\begin{eqnarray}
\alpha_0 N^{q-1} &=& \psi\, \alpha_1^p \alpha_0^{q-p-1} {q-1 \choose p}\,, \nonumber\\
\alpha_1 N^{q-1} &=& \psi\, \alpha_1^{p-1} \alpha_0^{q-p}  {q-1 \choose p-1}\,.
\label{c2}\end{eqnarray}
Solving the set of equations in (\ref{c2}) we obtain:
\bea
\alpha_1^{q-2} &&= \frac{N^{q-1} }
{\psi \left( \frac{q}{p} - 1\right)^{ (q-p)/2} {q-1 \choose p-1} }\,,\nonumber\\
\alpha_0^2 &&= \alpha_1^2\left( \frac{q}{p}-1 \right)\,,
\label{c3}
\eea
which gives:
\begin{equation} 
N=\alpha_0^2 + \alpha_1^2= \left(\frac{q}{p}\right)^{(p-1)/(q-2)}
 \frac{N^{ 2(q-1)/(q-2) } } 
 { \left (1- \frac{p}{q}\right)^{ (q-p)/(q-2) } 
 \left[ {q-1 \choose p-1} \right]^{1/(q-2)} }\,.
\label{c4}\end{equation}
Rearranging we obtain:
\begin{eqnarray}
N^q = \left(\frac{p}{q}\right)^{p-1} 
 \left (1- \frac{p}{q}\right)^{ q-p }  {q-1 \choose p-1} \,.
 \end{eqnarray}
\par
In the limit $q \gg p$ limit, we can approximate:
\begin{eqnarray}
&& {q-1 \choose p-1}  = \frac{ (q-1)!}{ (p-1)! (q-p)!} \approx \frac{1}{(p-1)!} q^{p-1} + 
 {\cal{O}}\left( q^{p-2} \right)\,,
 \nonumber\\
&&  \left (1- \frac{p}{q}\right)^{ q-p } \approx \exp(-p)\,,
 \end{eqnarray}
 which leads to:
 \begin{equation}
 \label{res4}
 N^q \approx \frac{p^{p-1} \exp(-p) }{(p-1)!} + {\cal O} \left( \frac{1}{q} \right)\,.
 \end{equation}
 
 \subsection{Results}
 We show below a graph of our results for the entanglement measure as a function of the number of qubits for the four cases solved in this section. 
 \par\begin{figure}[H]
\begin{center}
\includegraphics[width=10cm]{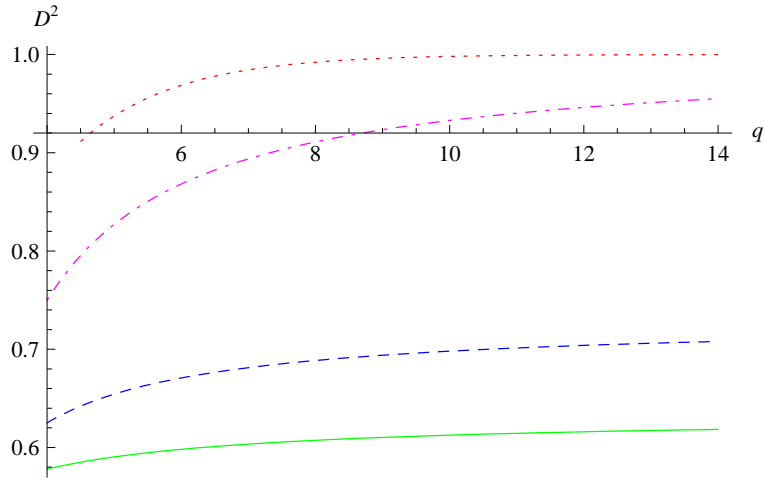}
\end{center}
\caption{The entanglement as a function of $q$ from Eq. (\ref{cdistance}) and Eqs. (\ref{res1}) - dotted/red, (\ref{res2}) - solid/green, (\ref{res3}) - dashed/blue and (\ref{res4}) - dot-dashed/magenta.}
\label{plot}
\end{figure}

\section{Conclusions}
\label{concSECT}
We have considered a generalisation of the usual
geometric measure of entanglement of pure states using the
distance to the nearest unnormalised product state. This definition does not
lead to any computational advantages, since the set of equations that 
determine the measure are still non--linear in general.  
However, our definition does provide 
an interpretation of the standard entanglement measure 
as the distance to the closest product state. We have  also found a relationship between the  norm and components of the closest separable state,
and the coefficients and basis states
of the Schmidt decomposition of the state $\ket{\psi}$.
For several cases where the target state has a large degree of symmetry, we have  solved the system of non--linear equations analytically, and looked specifically at the limit where the number of qubits is large. These results indicate that our new definition of entanglement, while similar to other definitions that can be found in the literature, is worthy of further study.

\acknowledgments
R. Kobes and G. Kunstatter gratefully acknowledge valuable discussions with Dylan Buhr and Dan Ryckman. 
This research was supported by the Natural Sciences and Engineering Research
Council of Canada.

\end{document}